\documentclass{jfm}
\usepackage{graphicx}
\usepackage{newtxtext}
\usepackage{newtxmath}
\usepackage{natbib}
\usepackage[dvipsnames]{xcolor}
\usepackage[colorlinks=true,linkcolor=MidnightBlue,urlcolor=MidnightBlue,citecolor=MidnightBlue,anchorcolor=MidnightBlue]{hyperref}
\usepackage{amsmath,graphicx, mathtools,multirow,gensymb, amssymb}
\usepackage{graphicx}
\usepackage{subcaption} 
\usepackage{algorithm}
\usepackage{algorithmic}
\usepackage{setspace}
\usepackage{makecell}
\usepackage{subcaption}
\usepackage{array}
\newcommand{\PreserveBackslash}[1]{\let\temp=\\#1\let\\=\temp}
\newcolumntype{C}[1]{>{\PreserveBackslash\centering}p{#1}}
\newcolumntype{R}[1]{>{\PreserveBackslash\raggedleft}p{#1}}
\newcolumntype{L}[1]{>{\PreserveBackslash\raggedright}p{#1}}

\newcommand{\RomanNumeralCaps}[1]

\title{Inferring activity from the flow field around active colloidal particles using deep learning}

\author{
Aditya Mohapatra\aff{1}, 
  \corresp{\email{adityamohapatra217@gmail.com}},
Aditya Kumar\aff{1}, 
Mayurakshi Deb\aff{1}, 
Siddharth Dhomkar\aff{1},
\and
Rajesh Singh\aff{1}
  \corresp{\email{rsingh@physics.iitm.ac.in}},
}
 \affiliation{\aff{1}Department of Physics, IIT Madras, Chennai 600036, India}

\begin{document}
\maketitle
\begin{abstract}
  Active colloidal particles create flow around them due to non-equilibrium process on their surfaces.  In this paper, we infer the activity of such colloidal particles from the flow field created by them via deep learning. We first explain our method for one active particle, inferring the  $2s$ mode (or the stresslet) and the $3t$ mode (or the source dipole) from the flow field data, along with the position and orientation of the particle. We then apply the method to a system of many active particles. We find excellent agreements between the predictions and the true values of activity. Our method presents a principled way to predict arbitrary activity from the flow field created by active particles. 
\end{abstract}

\section{Introduction}\label{sec:intro}
Active colloidal particles – such as microorganisms \citep{brennen1977,goldstein2015green}, and synthetic microswimmers \citep{ebbens2010pursuit, thutupalli2018FIPS} – are known to organise themselves in emergent structures via fluid-mediated interactions. An active particle can create flow around it even if it is not moving due to non-equilibrium process on its surface, which could be of biological origin \citep{brennen1977} or of phoretic origin \citep{ebbens2010pursuit}. Classifying the irreducible components of flow produced by active particles provides a handle through which we can tune inter-agent interactions to design higher order (hierarchical) assemblies of active particles which display a desired self-organization and tunable emergent dynamics. 

A topic of recent interest is the application of tools from machine learning to bear upon questions of inference and design principles in active matter systems \citep{cichos2020machine,tsang2020Learning,supekar2023learning,boffi2024deep}. Convolutional neural networks (CNNs) have been used for performing Particle Image Velocimetry \citep{Rabault2017pivcnn}, particle tracking\citep{tracking2016cnn} and feature extraction\citep{Hannel18feautrecnn}. To the best of our knowledge, a work on estimating activity from the flow field around active particles has not been reported. To this end, we use deep learning techniques given fluid flow around particles to infer the irreducible components of activity.

The remainder of the paper is organized as follows. In section \ref{sec:irredFLow}, we describe a method to compute irreducible components of flow around active particles. The methodology utilizing deep learning to learn activity from fluid flow around active particles is given in section \ref{sec:methodDL}. We apply the method to construct irreducible components of flow around an active particle in section \ref{sec:onePInf}. We extend this method to infer flow around many particles in section \ref{sec:manyPInf}.
 Finally, we conclude in section \ref{sec:conclusion} by summarizing our main results 
and suggesting directions for future investigations.

\section{Irreducible components of flow around active particles}\label{sec:irredFLow}
\begin{figure}
\centering
         \includegraphics[width=0.9\textwidth]{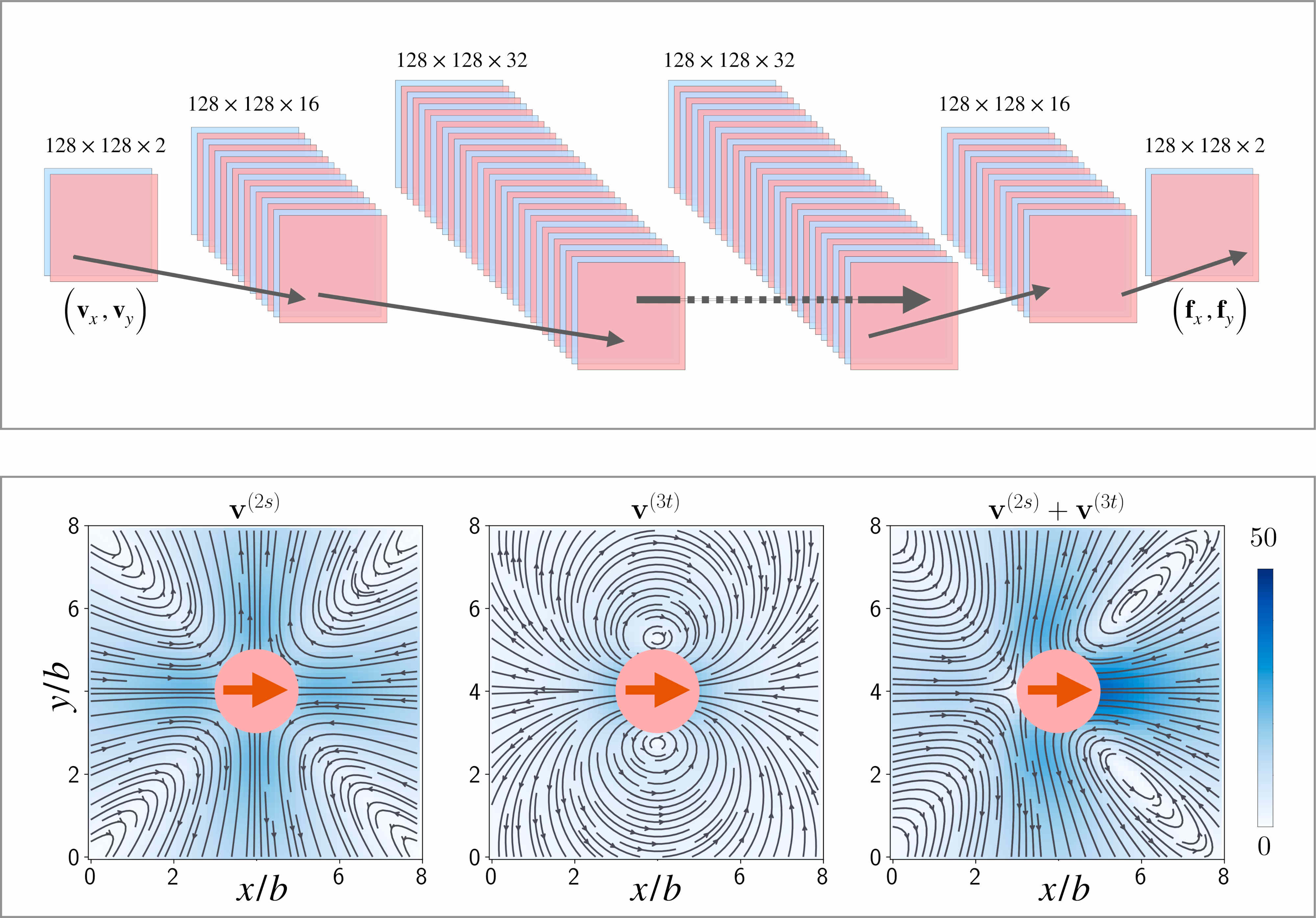}
    \caption{Top panel: the neural network architecture for computing force field from input velocity field. Each layer is Convolution layer with $3 \times 3$ kernel size and padding to preserve the dimension. Each block represents the image output of that layer in terms of $\text{Height} \times \text{Width} \times \text{Channels}$. The network goes till 128 channels (indicated by the dotted arrow).
    Bottom: Flow due to $2s$ and $3t$ modes of the active slip. A combined flow is also shown. Position $\mathbf c$ is shown by a circle, while the orientation $\mathbf p$ is shown by a solid arrow. 
    The flow streamlines have been overlaid on a
pseudo color plot of the speed.}
    \label{fig:compact-combined}
\end{figure}

In this section, we consider the problem of computing irreducible components of flow around active colloidal particles in a Stokesian fluid of viscosity $\eta$. 
At low Reynolds number, as appropriate to active colloidal particles, the fluid flow $\mathbf v$ satisfies the Stokes equation \citep{lauga2020fluid}
\begin{alignat}{1}
-\mathbf{\nabla}p +\eta\nabla^{2}\mathbf{v} & =-\mathbf{f},
\qquad 
\mathbf{\nabla}\cdot\mathbf{v}  =0.
\label{eq:stokes}
\end{alignat}
Here $p$ is the fluid pressure and $\mathbf f$ is the force density.
In appendix \ref{app:irredFLow}, we show that the flow around active particles can be written in terms of irreducible components $\mathbf{v}^{l\sigma}(\mathbf{k})$. The sum of the irreducible components in the Fourier space is:
\begin{align}
   {\mathbf{v}}^M(\mathbf{k}) = \sum_{l =1}^\infty  
   \sum_{\sigma\in\{s,a,t\}}
   \mathbf{v}^{l\sigma}(\mathbf{k}).
  \label{eq:IrredFLow}
\end{align}
As described in the appendix \ref{app:irredFLow}, the flow $\mathbf{v}^{l\sigma}(\mathbf r)$ decays as $r^{-l}$ in an unbounded three-dimensional flow. Here $\sigma$ can take three values for each $l$: $\sigma=s$, 
$\sigma=a$, and $\sigma=t$, which correspond to the three irreducible mode allowed for each $l$. See appendix \ref{app:irredFLow} for details. The terms $\mathbf v^{1s}$ is the flow due to a net external force, while $\mathbf v^{2a}$ is the flow due to a net external torque. Since an active particle has no external force or torque on it, the two leading irreducible components of exterior flow created by active 
particles are:
\begin{align}
   \mathbf{v}^{2s}(\mathbf{k})&=   \,\mathbf{G}(\mathbf{k})\cdot\mathbf f^{2s}, 
 \qquad 
 \mathbf f^{2s}=i\mathbf{k}\cdot\left[M\mathbf{Q}^{(2s)}\right]
   \\
   \mathbf{v}^{3t}(\mathbf{k})&=   \mathbf{G}(\mathbf{k})\cdot\mathbf f^{3t},
   \qquad \mathbf f^{3t}=-k^2 M\,\mathbf{Q}^{(3t)}. 
\end{align}
 The above two irreducible components of flow - the $2s$ mode (or the stresslet \citep{lauga2020fluid})
 and the $3t$ mode (or the source dipole \citep{lauga2020fluid}) - is known to fit flow around a wide range of active particles \citep{lauga2020fluid, marmol2024colloquium,goldstein2015green,ghoseIrreducibleRepresentationsOscillatory2014,delmotte2015large, drescherDirectMeasurementFlow2010,drescherFluidDynamicsNoise2011,baruah2025emergent,thutupalli2018FIPS, krishnamurthy2023emergent,guasto2010}. The leading fluid flow due to a force-free 
 and torque-free active particles is due to the two modes: $\mathbf{Q}^{(2s)}$ and $\mathbf{Q}^{(3t)}$. In addition, we can use uniaxial parameterization of the single layer density, which is true for a wide range of microorganisms having symmetry about a given axis. For such a scenario, with $\mathbf{p}$ being the orientation of the particle, we choose:   
 \begin{align}
      \mathbf{Q}^{(2s)} = s_0 \left( \mathbf{p} \mathbf{p} -\frac{1}{n_d} \mathbf{I}\right),\qquad 
      \mathbf{Q}^{(3t)} = d_0\,  \mathbf{p} .
 \end{align}
Here $n_d$ is the space dimension and $\mathbf{I}$ is the identity tensor, while $s_0$ and $d_0$ are constants. We note that the constants  $s_0$ and $d_0$ are related \citep{singh2015many} to the $B_1$ and $B_2$ modes of the squirmer model \citep{lighthillSquirmingMotionNearly1952,blake1971spherical,pedleySphericalSquirmersModels2016,lauga2020fluid}. 
The flow field due to these two modes is present in Fig. \ref{fig:compact-combined}.

For reasons explained above, we truncate the expression of the fluid flow in Eq.\eqref{eq:IrredFLow} to the leading modes  $2s$ (stresslet) and $3t$ (source dipole). 
For this choice, the expression for the net force density in the fluid, defined in Eq.\eqref{eq:stokes}, is written as the sum:
\begin{align}
 \mathbf{f}(\mathbf{r}) = 
 \mathbf f^{2s} + \mathbf f^{3t}=
 -
 \mathbf{\nabla} M(\mathbf{r})\cdot\mathbf{Q}^{(2s)}
 -
 {\nabla}^2 M(\mathbf{r})\,\mathbf{Q}^{(3t)}. 
\label{eq:FM_R} \end{align}
Here, $M$ is a Gaussian mollifier whose mean is at the center of the particle and the variance equals $2b^2/\pi$ \citep{maxey2001localized}, where $b$ is the radius of the particle. In the next section, we present a deep learning method to infer the individual components of active force density (and also the positions and the orientations of the particles) given a fluid flow data from the flow field around single and many active particles. See appendix \ref{app:irredFLow} for further details. 

\section{Methodology}\label{sec:methodDL}
Finding the velocity field solution to the Stokes equation with a given force density is discussed in the previous section. 
In addition, the velocity field may also be obtained from experimental systems using methods such as Particle Image Velocimetry (PIV)\citep{adrian1991piv,willert1991dpiv}. 
On the other hand, the inverse problem of finding the force density (and their irreducible components) from a 
given flow field data is non-trivial. This leads us to a neural-network-based learning approach for achieving the force density and corresponding irreducible modes using the universal function approximator capacity of neural networks \citep{hornik1989multilayerapprox}. 

To effectively approximate the inverse function using deep learning, we begin by constructing a sufficiently large and diverse dataset. This dataset is partitioned into training, validation, and test subsets. The model is trained on the training set to learn the mapping from velocity to force density. Throughout training, the validation set is used to monitor accuracy. Further the model is evaluated on the independent test set to verify the model's performance on unseen data. We consider flow data with both a fixed and variable number of particles. For a fixed number of particles, we consider a single particle in a grid and 4 particles in a grid, which can be generalized to $N$ particles. 
For a variable number of particles, we choose $N$ to vary from one to four. We generate datasets of 4000 realizations corresponding to systems with different particle counts on a $128 \times 128$ domain. The grid is divided into 
equal sub-regions, and one particle is randomly placed within each to ensure non-overlapping configurations in the case of more than one particle.

The $i$th particle, whose center is located at $\mathbf c^{(i)}=(x^{(i)}, y^{(i)})$, is assigned an orientation vector $\mathbf{p}^{(i)} = (\cos\theta^{(i)}, \sin\theta^{(i)})$, where $\theta^{(i)}$ is an angle made by the orientation vector with the $\hat x$-axis. The angle $\theta^{(i)}$ is selected from a discretized set of angles with $1^\circ$ resolution. Discretization of orientation space facilitates improved learning by the neural network. Each particle is assigned randomized stresslet strength $s_0 \in [800, 1200]$ and source dipole strength $d_0 \in [12000, 16000]$. The data are partitioned into training, validation, and test sets comprising 2800, 600, and 600 samples, respectively.

The neural network used in this work is a fully convolutional architecture designed to approximate the inverse  problem of Eq.\eqref{eq:stokes}. 
The input is a two-channel ($x$ and $y$ components of the fluid velocity $\mathbf v$) tensor \( \mathbf v \in \mathbb{R}^{H \times W \times 2} \), representing the force field on a two-dimensional grid of height \( H \) and width \( W \). The network output is \( \tilde{\mathbf f} \in \mathbb{R}^{H \times W \times 2} \), which approximates the {flow} field for the given velocity field. Each convolutional layer ${(l)}$ applies a localized linear transformation followed by a nonlinear activation function \citep{goodfellow2016deep}. The network consists of a composition of such layers as shown in fig. \ref{fig:compact-combined}. The operation at layer \( l \) is given by:
\begin{align}
 z^{(l+1)} = g(W^{(l)} * z^{(l)} + b^{(l)}) = h^{(l)} (z^{(l)})
 \qquad
 g(z) = \max(0, z)
\label{eq:NN} \end{align}
where \( z^{(l)} \in \mathbb{R}^{ H \times W \times C_l} \) is the input to the layer, \( W^{(l)} \in \mathbb{R}^{C_{l+1} \times C_l \times k \times k} \) is a learnable convolutional  kernel of size \( k \times k \), and \( b^{(l)} \in \mathbb{R}^{C_{l+1}} \) is a bias vector. Here, $C_l$ is the depth of $l$-th convolutional kernel.
The symbol \( * \) denotes the two-dimensional convolution operation. \( g \) is a nonlinear activation, ReLU,  defined in Eqn. \ref{eq:NN}. Here, ReLU is preferred over the sigmoid activation function because it helps prevent the saturation of artificial neurons\citep{pmlr-relu}. To improve generalization and reduce overfitting, dropout is applied after each activation with a dropout probability of \( 0.2 \). The neural network is trained with Mean Squared Error(MSE) as the loss function and Adam \citep{kingma2014adam} as optimizer. 

We note that our inference methodology is only applicable for low-Reynolds-number scenarios when the Stokes equation is a valid description of flow around particles. For active colloidal particles - such as microorganisms \citep{goldstein2015green} and synthetic microswimmers \citep{ebbens2010pursuit} - the Reynolds number is vanishingly small  and fluid flow is indeed Stokesian \citep{lauga2020fluid}.  Crucially, the existence of a valid Stokes flow implies that a one-to-one relation exists between the force density and fluid velocity \citep{lauga2020fluid}.
Thus, the inverse problem of obtaining the force density from a given Stokesian flow is well-posed for active colloidal particles. Indeed, we show that the CNN effectively learns a direct field-to-field mapping to obtain force density from flow field data for both single particle and multiple particles case accurately. {The method also works for data having variable number of particles. }
We can also infer about the contribution of stresslet and source dipole in the flow field using a new algorithm, as we describe later, that  
works for an arbitrary number of particles given a predicted force density from the neural network.

\begin{figure}
\centering
    \includegraphics[width=0.9\linewidth]{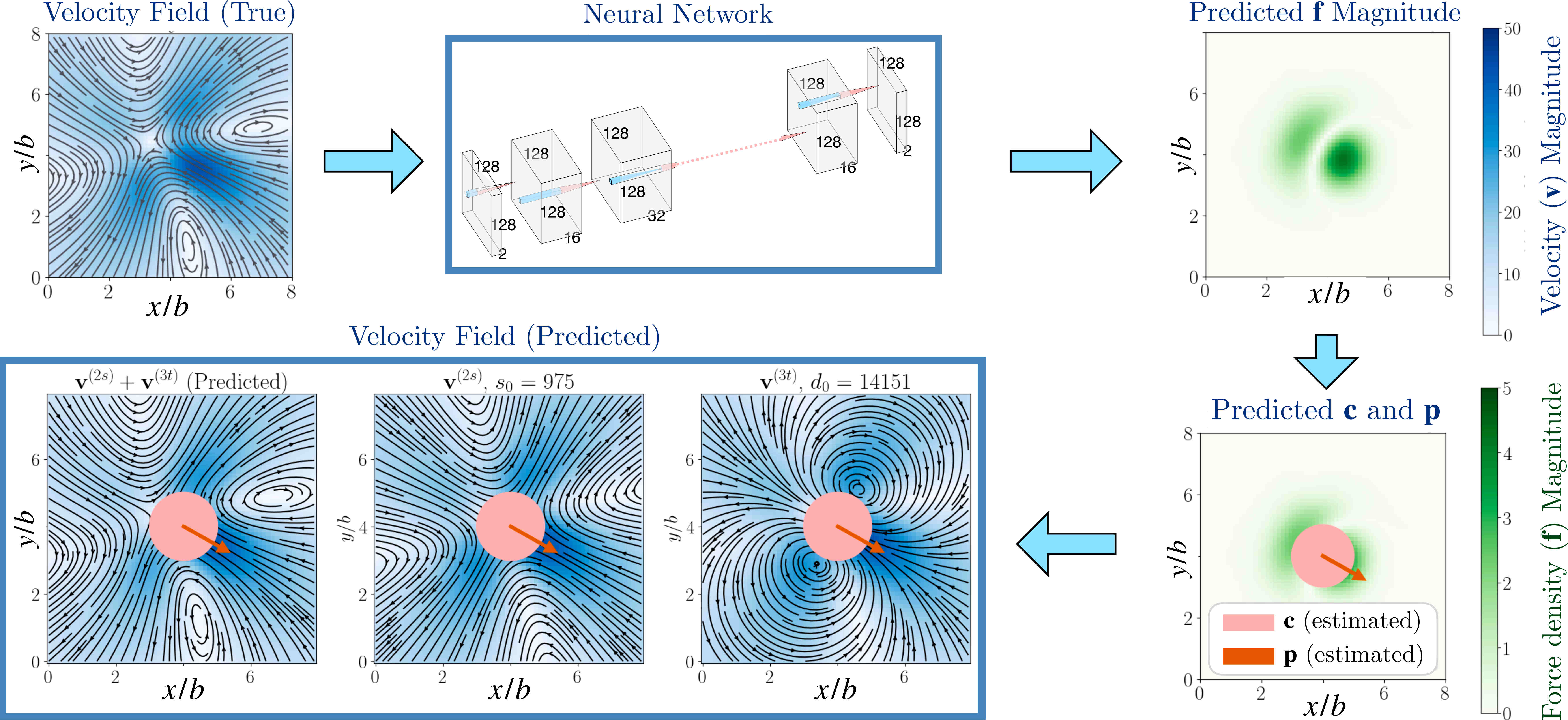}
    \caption{Flowchart of the method for the predictions of the model for a single particle: The velocity field for inference is shown along with a schematic of the neural network. The neural network predicts the force density field (here only magnitude is shown for clarity) from input velocity field. The velocity field obtained from predicted force is reconstructed with good accuracy. 
    See Algorithm \ref{alg} for details of the method.
    The flow streamlines have been overlaid on a
pseudo color plot of the speed.}
    \label{fig:1p_pred_force}
\end{figure}

\section{Inferring one particle flow}\label{sec:onePInf}
Active and self-propelled particles in the absence of applied forces or torques, with uniaxial parametrization, can be characterized by the flow generated.  The stresslet $\mathbf{Q}^{(2s)}$ generates flows that neither translate nor rotate the particle. The translational effect is generated by the source dipole $\mathbf{Q}^{(3t)}$. The sign of stresslet strength $s_0$ with respect to source dipole strength $d_0$ determines how the active particle acts. An active particle behaves as `puller', pulling the fluid inward along the swimming direction and forcing fluid away from their sides, if $s_0/d_0 > 0$. The swimmer is said to be a `pusher', pushing fluid away along their swimming direction and drawing fluid into the sides of their bodies, if $s_0/d_0 < 0$ \cite{lauga2020fluid}.

We consider the case of a single particle in a $128 \times 128$ grid, which is a slice from a large grid with periodic boundaries. The velocity field of the grid is fed into the neural network as input. The network computes the force density $\mathbf{\tilde f} = (\tilde f_x , \tilde f_y)$ by approximating the inverse problem accurately as shown in \ref{fig:1p_pred_force}. The orientation $(\mathbf{p})$ of the particle is computed by taking local average of predicted force density in the neighborhood where the magnitude of predicted force density is maximum. The position  $(\mathbf{c})$ of the particle, strength of stresslet ($s_0$) and the source dipole  ($d_0$) are estimated by minimizing the squared error of true force density and estimated force density with gradient descent. The algorithm for estimating these parameters is given in Algorithm \ref{alg}. 
\begin{algorithm}
\caption{: Particle Detection and Parameter Estimation for $N$ Particles}
\label{alg}
\begin{spacing}{1} 
\begin{algorithmic}[1]
\REQUIRE Force density $\tilde f_x$, $\tilde f_y$; threshold ratio $\tau$; particle size $b$; minimum distance $d_{\text{min}}$

\STATE Compute force density magnitude: 
$
\tilde f_{\text{mag}}(x, y) \gets \sqrt{\tilde f_x^2(x, y) + \tilde f_y^2(x, y)}
$

\STATE Obtain global maximum: $\tilde f^*_{\text{mag}} \gets \displaystyle \max_{x,y} \tilde f_{\text{mag}}(x, y) $

\STATE Detect peaks: 
\[
\mathcal{P} \gets \left\{(x, y) : (x,y) \notin \mathbf{c}^{(i)} \pm d_{\text{min}} \ \text{and} \ \tilde f_{\text{mag}}(x, y) > \tau \cdot \tilde f^*_{\text{mag}} \right\}
\]

\STATE Detect number of particles  $N \gets ||\mathcal{P}||$ and compute center of mass $\{\mathbf{c}^{(i)}\}_{i=1}^{N}$ for each region of $2d_{min} \times 2d_{min}$ centered around peak.

\FOR{each $\mathbf{c}^{(i)}$}
    \STATE Initialize $s_0^{(i)}, d_0^{(i)}$
    \STATE Estimate local direction: 
    $
    \mathbf{p}^{(i)} \gets (\bar{\tilde f}_x, \bar{\tilde f}_y) \quad \text{in neighborhood of } \mathbf{c}^{(i)}
    $
    
    \STATE Initialize best loss: $\mathcal{L}_{\min} \gets \infty$

    \FOR{$\mathbf{c}^{(i)} \gets \mathbf{c}^{(i)} + \Delta \mathbf{c}$}
        \STATE Define estimated force field using Gaussian envelope $M$: 
        \[
            \mathbf{\hat f} = s_0^{(i)} \nabla  M(\mathbf{c}^{(i)}) \cdot \left( \mathbf{p}^{(i)}\mathbf{p}^{(i)} - \frac{\mathbf{I}}{2}\right)
                - d_0^{(i)} \nabla^2 M\left(\mathbf{c}^{(i)} \right)  \mathbf{p}^{(i)}
        \]

        \STATE Minimize loss over the domain $\Omega$: 
        \[
        \mathcal{L}  = \sum_{\Omega} 
        \left\| 
        \,\mathbf{\tilde f} - \mathbf{\hat f} \,
        \right\|^2
        \]

        \STATE Update parameters if $\mathcal{L} < \mathcal{L}_{\min}$
    \ENDFOR

    \STATE Store $(s_0^{(i)}, d_0^{(i)}, \mathbf{p}^{(i)}, \mathbf{c}^{(i)})$
\ENDFOR

\RETURN $\{(s_0^{(i)}, d_0^{(i)}, \mathbf{p}^{(i)}, \mathbf{c}^{(i)})\}_{i=1}^{N}$
\end{algorithmic}
\end{spacing}
\end{algorithm}

\begin{figure}
\centering
    \includegraphics[width=0.984\textwidth]{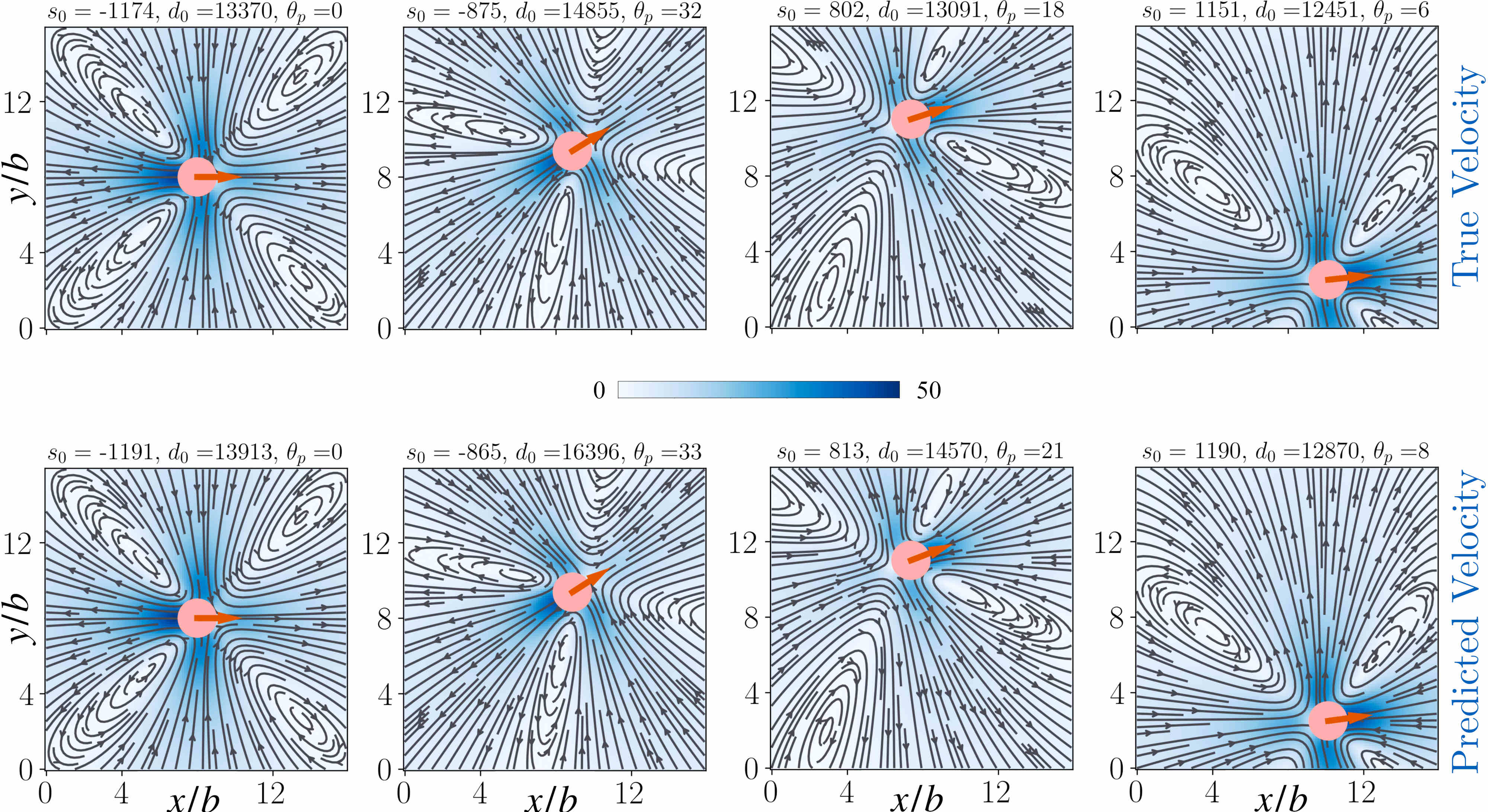}
    \caption{True (top row) and predicted (bottom row) velocity fields due to different values of stresslet strength $(s_0)$, source dipole strength $(d_0)$, orientation $(\mathbf{p})$ and particle position $(\mathbf{c})$. Values have been rounded to the nearest integer for clarity.  We show four distinct examples for illustration.
    The flow velocity streamlines have been overlaid on a pseudo color plot of the flow speed.
    Here, $\theta_p$ is the angle made by the particle with the positive $x$-axis of the Cartesian coordinates.
    }
    \label{fig:var_sd}
\end{figure}

The prediction of the neural network gives a very accurate representation of force density when compared with the ground truth as seen in figure \ref{fig:1p_pred_force}. We reconstruct the velocity field using the parameters obtained from the predicted density using Algorithm \ref{alg}. %
The algorithm 
returns position and orientation of the particle along with the strengths of the stresslet  ($2s$ mode) and source dipole ($3t$ mode).
These parameters can be used to reconstruct velocity field  using Equations \eqref{eq:stokes}-\eqref{eq:FM_R}. The predicted velocity field represents the flow with a good accuracy and we are able to decompose the flow into $2s$ and $3t$ modes.
The predictions of the stresslet and source dipole strength give very close result with respect to the ground truth having $3.26 \%$ and $5.70\%$ error respectively averaged over 600 test cases unseen to the neural network. We have predicted the parameters using the neural networks for particles with arbitrary positions and orientations and reconstructed the velocity field (see Figure \ref{fig:var_sd}). 

\section{Inferring many particles flow}\label{sec:manyPInf}
We extend our model further to test with flow due to $N$ particles. Here we consider the case of $N=4$. In an arbitrary velocity field, we first detect the number of particles in the box, detecting the number of locations where force density magnitude is concentrated over some small local area using scikit-ndimage \citep{van2014scikit}. We ensure there is sufficient distance between two different regions associated with particles while detecting. The approximate particle position can be guessed as the center of mass (force density magnitude) of that region. Now each of these regions can be thought as a single particle system and the estimation methods of single particle can be applied. The process is explained in Algorithm \ref{alg}. 

\begin{figure} 
\centering
    \includegraphics[width=\linewidth]{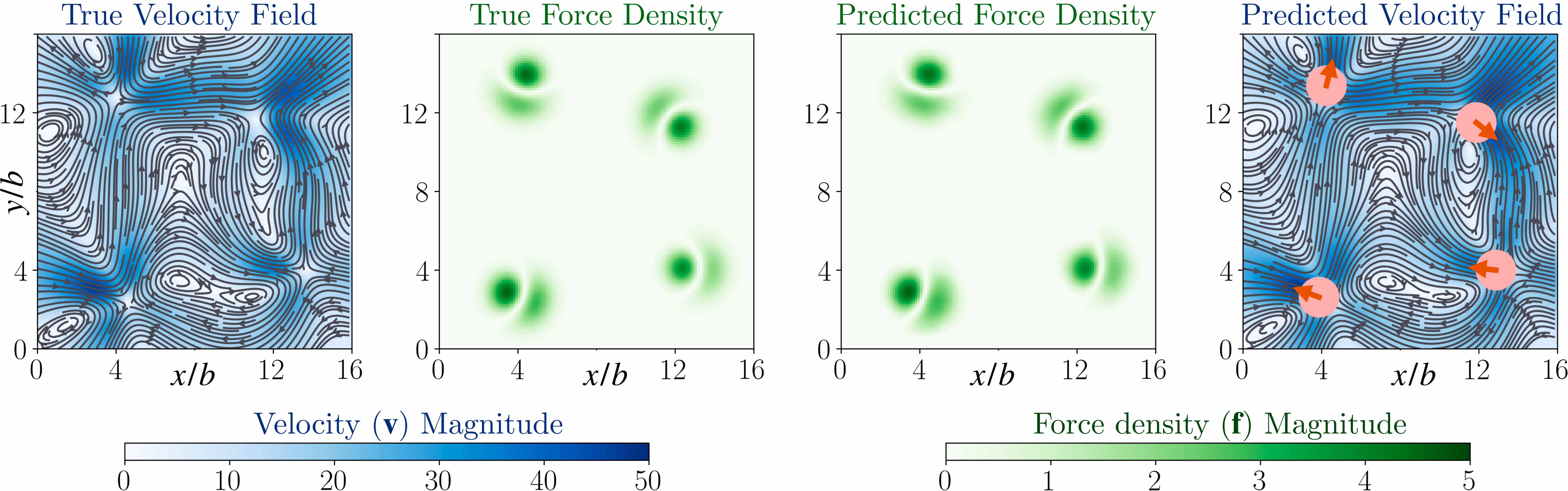}
    \caption{Predictions of the model for multiple particles in a grid.
        The flow velocity streamlines have been overlaid on a pseudo color plot of the flow speed. Magnitude of the force density is also shown. Our inference method using deep learning predicts location and orientation of the particles as well as the force density in the fluid due to them.}
    \label{fig:4p_force_pred}
\end{figure}

The neural network is able to predict the force density effectively as seen from the comparison with ground truth force density in Fig. \ref{fig:4p_force_pred}. From the predicted force density we get the parameters by Algorithm \ref{alg}. {The final output of this procedure is a set of inferred parameters for all particles: $\{\mathbf p^{(i)}$, $\mathbf c^{(i)}$ $s_0^{(i)}$, and $d_0^{(i)}\}_{i=1}^N$.
These parameters can be used to reconstruct the velocity field using the results of section \ref{sec:irredFLow}.}
Thus, our method using deep learning to extract force density and then fitting that data to the known model for the fluid flow allows prediction in a system of many particles. 

\begin{figure} 
    \centering
    \includegraphics[width=\linewidth]{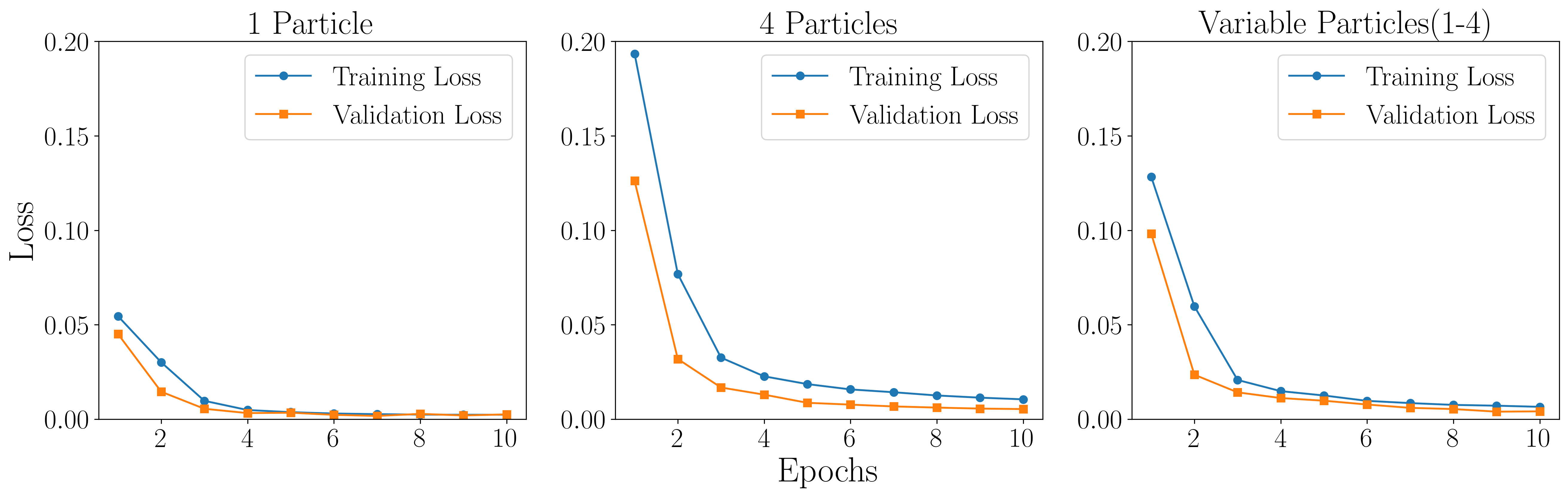}
    \caption{
    {The training loss and validation loss over epochs indicating that the model is learning well
without overfitting. We show this for cases with fixed number of particles ($N=1$ and $N=4$ are shown) and variable number of particles ($N$ varies from 1 to 4). }
    \label{fig:lossF}
    }
\end{figure}

In addition, our method can be applied to a variable number of particles. 
We note that the our network is capable of predicting the force density accurately even if the number of particles is varied, while the Algorithm \ref{alg} is agnostic to the number of particles. 
We plot the training loss and validation loss in Figure \ref{fig:lossF} for cases of fixed and variable number of particles. For the case of a fixed number, we show data for system with one and four particles, while in the case of a variable number of particles, we have considered one to four particles. Histogram of error is shown in  Fig.\ref{fig:lossHist} of the appendix. 
Thus, robustness of the method is demonstrated by showing that the model works for a variable number of particles. Moreover, we show that the model works even when we change spatial resolution of the data. See Fig.\ref{fig:res} in appendix \ref{app:quant}. 
\section{Summary and Conclusions}\label{sec:conclusion}
In this paper, we have studied the inverse problem of computing activity of colloidal particles from the specified flow field data around them by combining artificial intelligence and physics-based methods. Such a method is particularly attractive since PIV data from experiments can provide us with flow field data. 
Using deep learning, we predict activity from the two-dimensional flow field data created by active particles. We achieve this by inverting the flow field data to obtain force density in the fluid using CNNs. The reconstructed flow field from the predicted force density is in excellent agreement with the true flow field. Finally, using a new inference algorithm, we infer the position and orientation of active particles along with irreducible components of the velocity (stresslet and source dipole).

Extending our method to infer the arbitrary modes of active slip around many active particles for a given flow data in two- and three-dimensional systems suggests an exciting direction for future work. In this paper, we have used deterministic fluid velocities around active particles in two-dimensions to infer their activity. Another potential future direction is to consider cases when there is noise in the fluid velocity by denoising and reconstruction methods \citep{weng2001wavelet,wang2022dense}. 
Comparison of different architectures to find the most optimal one for these applications is another direction for future work. For generality, we have used a simple loss function in our CNN architecture. Using a physics informed neural network with a more problem-specific choice of loss function suggests another exciting avenue for future work.

{In this paper, we have used CNNs to solve an inverse problem to obtain force density given fluid flow data. 
This inverse problem has inherent non-trivial spatial correlations.
The CNNs are particularly well-suited for solving inverse problems involving spatial correlations because they are inherently designed to capture and learn hierarchical spatial features through convolutional filters \citep{goodfellow2016deep}. 
{In addition, we show in  Fig.\ref{fig:ablataion} and appendix \ref{app:quant} that the ability of CNN to learn non-linear mapping is crucial for our case by training the convolution layers with and without activation functions.}
Moreover, CNNs are desirable for large datasets \citep{goodfellow2016deep}, which is the case for inference from flow field data, especially when larger numbers of particles are considered.
We also note that the methodology using CNNs presented in this paper is only expected to work for Stokesian fluid flow data. For more complex settings, when the steady Stokes equation, Eq.\eqref{eq:stokes}, is no longer a valid description of fluid flow - such as scenarios where inertia and/or memory effects in the fluid are important -  entirely different
architectures and strategies of training may be needed.}
\\\par

\noindent
\textbf{Declaration of Interests}. The authors report no conflict of interest.
\\\par

\noindent
\textbf{Code availability}. All code used in this work is freely available on GitHub (https://github.com/soft-matter-physics/deep-learning-active-flows). 
\appendix
\section{Solving the Stokes equation to compute flow around active particles} \label{app:irredFLow}
The flow around a finite sized rigid sphere of radius $b$ in a fluid of viscosity $\eta$ with single layer density $\mathbf q$ is given by the boundary integral expression \citep{pozrikidis1992}:
\begin{align}
  \mathbf{v}(\mathbf{r}) = -\int \mathbf{G}(\mathbf{r} - \mathbf{r}^{\prime}) \cdot \mathbf{q}(\mathbf{r}^{\prime}) dS 
  \label{eq:BIE}
\end{align}
Here $\mathbf r$ is the point at which the fluid velocity is computed. $\mathbf G$ is a Green’s function of Stokes equation and $S$ it the surface of the particle, while $\mathbf r’$ is an integration variable over the surface. 
The above integrals can be solved by expanding the boundary fields in TSH (tensorial spherical harmonics) \citep{hess2015tensors}.
The TSH are defined as
$    Y_{\alpha_{1}\dots\alpha_{l}}^{(l)}(\hat{\mathbf{b}}) 
    =(-1)^{l}\,b^{(l+1)}\,\nabla_{\alpha_{1}}\dots\nabla_{\alpha_{l}}\frac{1}{b}$.
    Here $\hat{\mathbf b} = \mathbf b/b$, while $b=||\mathbf b||$ and $\mathbf b$ is the radius vector of the active colloidal sphere.
The expansion of the single layer density $\mathbf{q}$ in TSH is:
\begin{subequations}
\begin{align}
\mathbf{q} (\mathbf{b} ) & =\sum_{l=1}^{\infty}w_{l}\,\mathbf{Q} ^{(l)}\cdot\mathbf{Y}^{(l-1)}(\mathbf{\hat{\mathbf b}} ),\qquad
\mathbf{Q} ^{(l)}  =\tilde w_{l}\int \mathbf{q}
 \,\mathbf{Y}^{(l-1)}(\mathbf{\hat{\mathbf b}} ),
\end{align}
\end{subequations}
Here, the $l-$dependent constants are: ${w}_{l}=\frac{2l-1}{4\pi b^{2}}$ and 
$\tilde w_{l}=\frac{1}{(l-1)!(2l-3)!!}$ 
\citep{hess2015tensors}. 
In the above, and throughout the paper, a maximal contraction of
two tensors is denoted by a dot product. 
The boundary integral of Eq.\eqref{eq:BIE} can be solved exactly as follows: (a) expanding the boundary fields in TSH,
(b) performing a Taylor expansion of kernels (the Green's function) about the center of the particles, and 
(c) using the orthogonality of TSH \citep{singh2015many,singh2018microhydrodynamics}.
The final result is the following summation:
\begin{align}
\label{eq:flowQL}
  \mathbf{v}(\mathbf{r}) = -\sum_{l=1}^{\infty} \left(
  1 + \frac{b^2}{4l+2}\nabla^2 \right)
  \mathbf{\nabla}^{(l-1)}\mathbf{G}(\mathbf{r})\cdot\mathbf{Q}^{(l)},  
 \end{align}
Thus, the flow due to $l$th mode decays as $\nabla^{l-1}\mathbf G(\mathbf r)$. In three-dimensions $\mathbf G(\mathbf r)\propto r^{-1}$, and thus, the flow due to any of the $l\sigma$ mode as $r^{-l}$ \citep{singh2015many,deb2025ewald}, where $r$ is the distance from the source. 
The equation \eqref{eq:stokes} in the Fourier space is:
\begin{align}
    \mathbf {v}(\mathbf{k}) & =
\mathbf {G}
(\mathbf{k})\cdot
\mathbf {f}(\mathbf{k}),\quad\quad
\mathbf {G}(\mathbf{k})=\frac{1}{\eta k^2  }\left({\mathbf{I}} -\frac{\mathbf{k}\mathbf{k}}{k^{2}}\right). 
\label{eq:stokes-kspace}
\end{align}
Here, $\mathbf I$ is the identity tensor and $\mathbf k$ is Fourier wave vector. 
We define Fourier transform of a function $\varphi(\mathbf{r})$ and Dirac delta function $\delta (\mathbf r)$ as
\begin{align}
{\varphi}(\mathbf{k})=
\int\varphi(\mathbf{r})\,e^{-i\mathbf{k}\cdot\mathbf{r}}\,d\mathbf{r},\qquad
\delta(\mathbf{r})=\frac{1}{(2\pi)^{3}}\int  e^{i\mathbf{k}\cdot\mathbf{r}}\,d\mathbf{k}.
\label{eq:FT} 
\end{align}
Using equations \eqref{eq:stokes-kspace} and \eqref{eq:flowQL}, we can identify an effective force density:
\begin{align}
 \mathbf{  f} (\mathbf{k}) = -\sum_{l=0}^{\infty} \left(1 - \frac{b^2}{4l+2}k^2 \right)(i\mathbf{k})^{(l-1)}\cdot\mathbf{Q}^{(l)}
 \end{align}
that produces the flow in Eq. \eqref{eq:flowQL}. Using the definition of Dirac delta function, the above force density can be written in the coordinate space as:
\begin{align}
 \mathbf{f}(\mathbf{r}) =- \sum_{l=1}^{\infty} \left(1 + \frac{b^2}{4l+2}\nabla^2 \right)
 \mathbf{\nabla}^{l-1}
 \delta(\mathbf{r})\cdot\mathbf{Q}^{(l)}. 
 \end{align}
that satisfies the Stokes equation. Since our interest is in the external flow, we are free for this purpose to replace the $\delta$ function in the above expression with a mollifier $M$ which produces the same exterior flow \citep{maxey2001localized,delmotte2015large,singh2018microhydrodynamics}. We choose a Gaussian mollifier whose mean is at the center of the particle and the variance equals $2b^2/\pi$ \citep{maxey2001localized}. Using Eq.\eqref{eq:stokes-kspace}, the expression of the flow in Fourier space is 
\begin{align}
 { \mathbf{v}}^M(\mathbf{k}) &= -\sum_{l=1}^{\infty} 
 \left(1 - \frac{b^2}{4l+2}k^2\right)
 (-i\mathbf{\mathbf k})^{l-1}\, M(\mathbf{k})\mathbf{G}(\mathbf{k})\cdot\mathbf{Q}^{(l)} 
 \label{eq:VMF}\end{align}
The coefficients of single layer density
 $\mathbf{Q} ^{(l)}$ 
are $l$-th rank tensors, symmetric irreducible in their last $l-1$
indices. 
The reducible tensors $\mathbf{Q} ^{(l)}$  
can be written in irreducible forms as:
$\mathbf{Q} ^{(l\sigma)}$, where $\sigma=s$ represents 
symmetric-traceless part of the tensor, while $\sigma=a$ represents the antisymmetric part, and $\sigma=t$ denotes the trace. 
Here, $\mathbf{Q} ^{(l\sigma)}$
are irreducible tensors of rank $l$, $l-1$ and $l-2$ for $\sigma=s,a,$
and $t$ respectively. 
The final expression of the flow - Eq.\eqref{eq:IrredFLow} of the main text - in terms of irreducible components is obtained when Eq.\eqref{eq:VMF} is written as a sum over
the irreducible components of $\mathbf Q^{(l)}$.

\begin{figure} 
    \centering
    \includegraphics[width=0.758\linewidth]{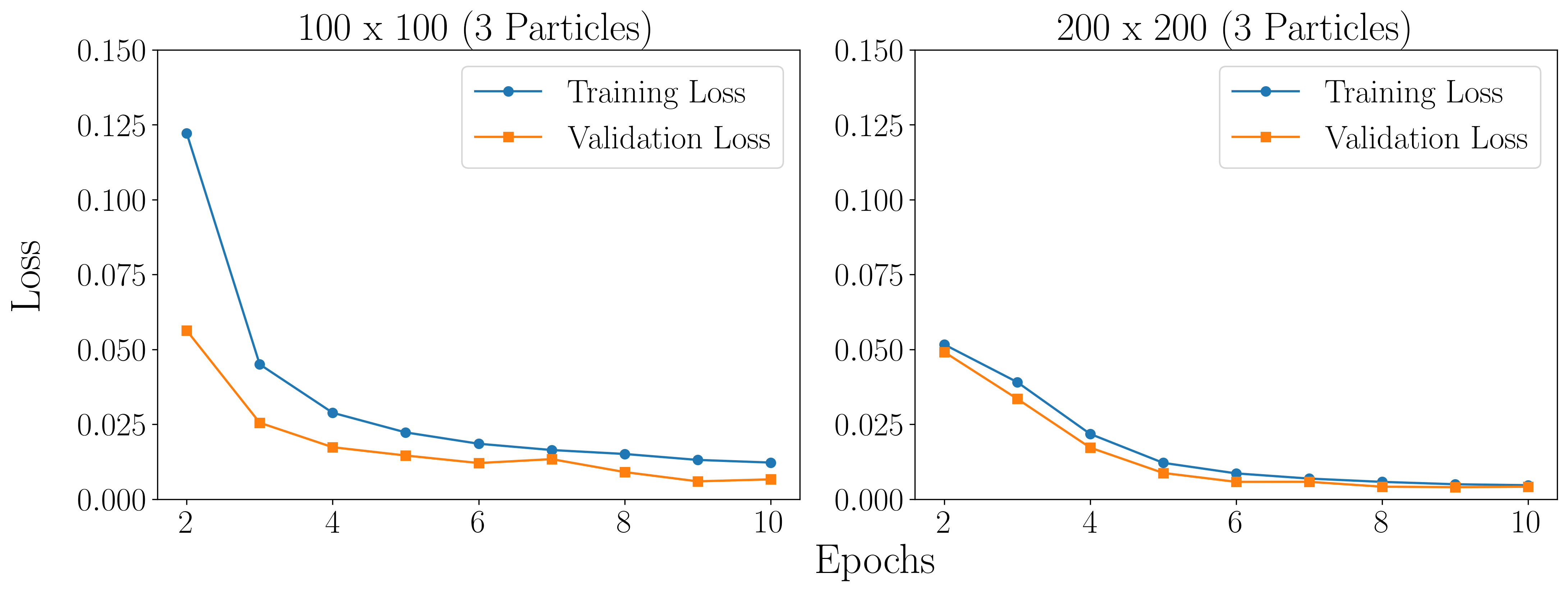}
    \caption{ {Loss function as a function of epochs for two different resolutions of the data.}}
    \label{fig:res}
\end{figure}

{\section{Quantitative Analysis\label{app:quant}}
\subsection{Root Mean Square Error}
The convergence plot (Average of Mean Squared Error over batches of 32 vs Epoch) is given in Fig.\ref{fig:lossF} of the main text. It shows the good fit in neural network. The network was tested over 600 test cases in each of the single particle, multiple particle and variable particle instances. The RMSE(Root Mean Squared Error) of the true force density versus the one predicted by the network over the cases is plotted for each of the instances as histogram Fig.\ref{fig:lossHist}. The RMSE is calculated over all grid points for a single instance.

{\subsection{Variable Resolution}
In Fig.\ref{fig:res}, we demonstrate that our network works for variable spatial resolution. In the main text, we have set the spatial resolution to be $128\times 128$ while we vary number of particles. Here, we plot the loss function as a function of epochs for two different resolutions ($100\times 100$ and $200\times 200$) for an inference study of three particles. }
\begin{figure} 
    \centering
    \includegraphics[width=0.94\linewidth]{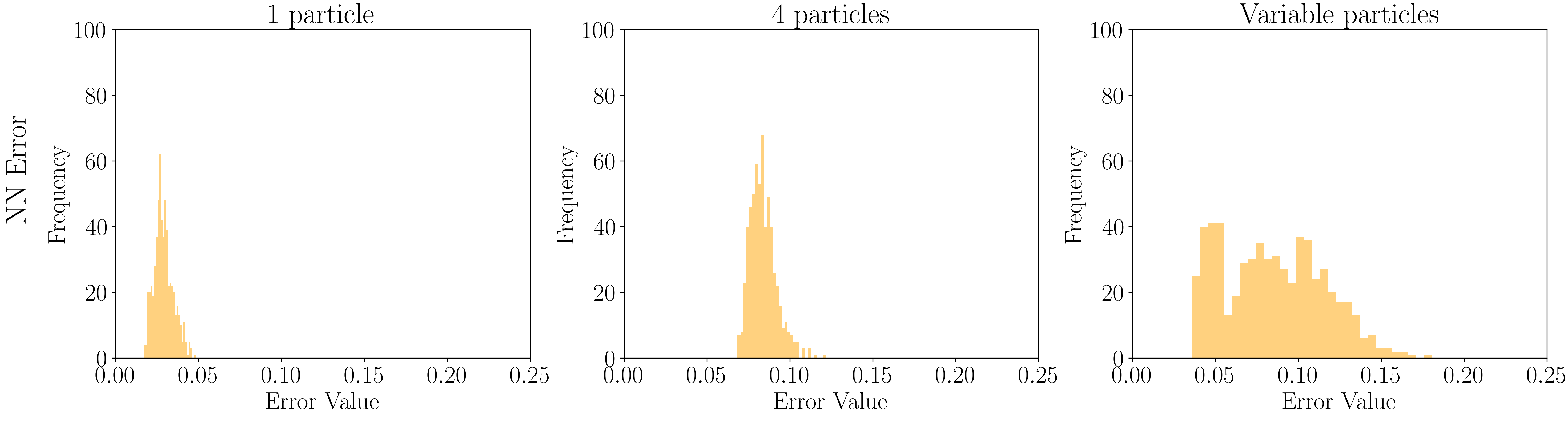}
    \caption{{Histogram of root mean squared error from the predictions by CNN.}}
    \label{fig:lossHist}
\end{figure}}

\begin{figure}
    \centering
    \includegraphics[width=0.75\linewidth]{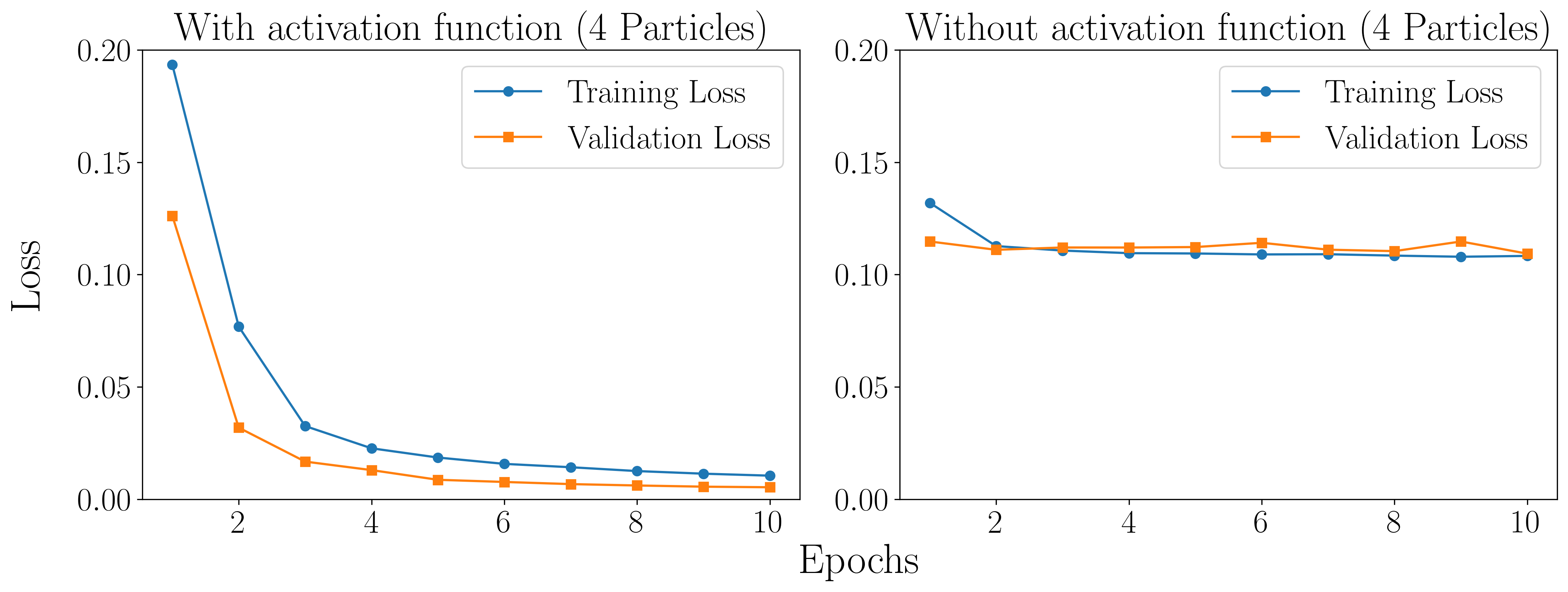}
    \caption{{Ablation study. Measuring loss with epochs using CNN with activation functions ({LEFT}) and 
    without activation function (RIGHT).}}
    \label{fig:ablataion}
\end{figure}

{\subsection{Ablation Study}\label{app:ablation}
The advantages of CNN in this particular problem is studied by checking the solutions without the components of CNN. To evaluate it, the dataset was trained with just convolution layers without any activation functions, which essentially makes it a linear map. The comparison of this case against the CNN with activation function is shown by the convergence plot in Fig.\ref{fig:ablataion}. The former is not converging with similar iterations where as the later converges rapidly.
{CNNs learn non-linear mappings using activation functions. Thus, we demonstrate the importance of non-linear mappings in deep learning by showing that CNN with activation function is essential for the model presented here to be effectively trained against data.}
}

\end{document}